# Stochastic models of population extinction

*Otso Ovaskainen[1] and Baruch Meerson[2]*

[1]Metapopulation Research Group, Department of Biosciences, University of Helsinki, PO Box 65 (Viikinkaari 1), FI-0014, Finland, email: otso.ovaskainen@helsinki.fi

[2]Racah Institute of Physics, Hebrew University of Jerusalem, Jerusalem 91904, Israel, email: meerson@cc.huji.ac.il

# Abstract

Theoretical ecologists have long sought to understand how the persistence of populations depends on biotic and abiotic factors. Classical work showed that demographic stochasticity causes the mean time to extinction to increase exponentially with population size, whereas variation in environmental conditions can lead to a power law scaling. Recent work has focused especially on the influence of the autocorrelation structure ('color') of environmental noise. In theoretical physics, there is a burst of research activity in analyzing large fluctuations in stochastic population dynamics. This research provides powerful tools for determining extinction times and characterizing the pathway to extinction. It yields, therefore, sharp insights into extinction processes and has great potential for further applications in theoretical biology.

# The importance and challenge of understanding population extinction

One of the most fundamental questions in population biology concerns the persistence of species and populations, or conversely their risk of extinction. Extinction risk is influenced by a myriad of factors, including interaction between species traits and various stochastic processes leading to fluctuations and declines in population size [1-7]. Assessment of extinction risk is necessarily scale-dependent [8, 9]: for example, a metapopulation might persist in a balance between local extinctions and re-colonizations, even though none of the local populations would persist alone [10-12]. Modeling approaches for quantifying extinction, such as population viability analyses, are often faced with so many levels of uncertainty that their utility has been questioned by some researchers [13, 14]. However, others have argued that models can be meaningfully used to quantify extinction risks [15, 16], and that the predicted extinction risk can be a more objective measure for classifying species as red-listed than many other metrics [17, 18].

Conducting a population viability analysis involves the steps of choosing an appropriate model, fitting the model to data, and using the fitted model to predict the extinction risk [15]. In this review, we consider the last step only, asking how predicted extinction risk depends on the structure and parameters of a stochastic population model. Motivated by the material summarized in Box 1, we focus especially on studies that have derived explicit formulae for the mean time to extinction (MTE). We further narrow down our scope to the scale of a single local population consisting of one (or few) species. While extinction risk can be greatly influenced by spatial structure [10-12, 19-24] and the community context [25], understanding the fate of a single local population is the fundamental building block that is also required in the analyses of more complex models [10, 26].

We start by briefly reviewing the classical results on extinction risk in the context of stochastic population modeling. We then comment on the mathematical methods available for extinction studies, especially highlighting the methodological advances in the physics literature. We cover a sequence of biologically relevant models, starting from single species models in a stable environment, and then extending this to include environmental variability, Allee effects, and models with multiple entity types.

## Classical results

To place the recent developments into a perspective, we start by reviewing the highly influential works of Leigh [27], Lande [28] and Foley [29]. These authors showed, for different variants of the canonical model (Box 2), that in a stable environment the MTE of a local population grows exponentially with the carrying capacity $K$ [27, 28],

$$\text{MTE} = C_1 \exp(bK), \quad (\text{Eq. 1})$$

whereas under sufficiently strong, uncorrelated environmental stochasticity, the dependence is power law [27-29],

$$\text{MTE} = C_2 K^c. \quad (\text{Eq. 2})$$

The parameters $b > 0$ and $c > 0$ determine, to the leading order, how MTE depends on carrying capacity $K$. The parameters $C_1$ and $C_2$ (called pre-factors) can also depend on $K$, but their relative role becomes unimportant when the carrying capacity is large. In his pioneering work, Leigh [27] modeled demographic stochasticity by considering a birth and death process in a finite population (Box 1). Assuming a self-regulating population, Leigh ended up with the exponential scaling law (Eq. 1) with parameter $b = r/v_d$, where $r$ denotes the population growth rate and $v_d$ is the variance of the growth rate due to demographic stochasticity. Lande [28] used a continuous variable for population size, and assumed that the population grows without regulation until it reaches the carrying capacity, at which point the growth rate becomes zero. He arrived at the result $b = 2r/v_d$, where $v_d$ includes not only the variance due to demographic stochasticity but also the variance due to demographic heterogeneity [5] (i.e. variation in the birth and death rates among individuals).

All three papers [27-29] considered the impact of environmental stochasticity by assuming that the growth rate varies randomly over time with variance $v_e$, and showed that the power-law scaling (Eq. 2) holds with $c = 2r/v_e$ (we have reformulated some of the results to account for different interpretations of environmental noise [26]). In summary, the classical papers [27-29] showed that environmental stochasticity can lead to a substantial extinction risk also for large populations, not just small ones, and especially so if the population growth rate is low. The studies [27-29] have been very influential because their results were summarized into simple mathematical formulae (Box 1), and because the qualitative results are independent of many details of model structure, such as the exact form of the density dependence.

# Methods for estimating time to extinction

The process of extinction is difficult to study mathematically, and consequently most results in this area are approximations. It has recently become apparent that the approximations adopted in the classical papers [27-29] can give wrong answers for large parts of the parameter space. As the choice of the method can critically influence the results, we next review the methods available for estimating MTE in stochastic models of population dynamics.

To start with, given almost any kind of stochastic population model, Monte-Carlo simulations (Fig I in Box 2) can be used to characterize any aspect of population dynamics such as MTE. This approach is widely applied in the ecological literature [30-35] because of its ease of implementation. A serious disadvantage of this approach is that the results come in a numeric format, and so they can be difficult to interpret and generalize.

The first step towards an analytical insight is to write down the master equation corresponding to the stochastic model (Box 3), and solve it numerically [36, 37]. This approach can be used mostly in models with a single or two entity types but it is not feasible for models with many entity types, as the size of the state-space becomes prohibitively large. This approach avoids the Monte-Carlo error associated with simulations, but shares the drawback of the results being in numerical format.

In addition to providing numerical solutions, the master equation is a natural starting point for deriving analytical results. For any single-step model of a single entity type inhabiting a stable environment (Box 2), it is possible to derive an exact formula for the MTE [38]. Unfortunately, this formula is too cumbersome to give any analytical insight. However, for models in which density dependence either monotonically increases the per-capita death rate or decreases the per-capita birth rate, the exact formula can be simplified to yield a large-$K$ approximation [39], i.e. an approximation that becomes asymptotically exact as the carrying capacity $K$ increases (see Box 2 for an example). The method of [39] generalizes many of the earlier approaches applied separately for specific single-step models [40-44]. Although in some cases it misses pre-exponential factors in the MTE [45], the most important exponential dependence is described correctly.

A different approach is needed for models that incorporate additional features, such as variation in environmental conditions, Allee effects, multiple simultaneous births or deaths, or multiple entity types. In the ecological literature, the most widely used mathematical method for these types of problems is the Fokker-Planck (FP) approximation (also called the diffusion approximation) [27-29, 46, 47], that replaces the original master equation by a partial differential equation or an equivalent stochastic differential equation called the Langevin equation [38] (Box 3), and the related moment-closure approximation [44]. These approximations can be used to derive a formula for the MTE [27-29, 46, 48] and to fit population models to time-series data [29, 46-50].

Unfortunately, the FP approximation in general fails to correctly describe large fluctuations, such as those leading to population extinction [36, 39, 51]. The FP approximation faithfully describes typical population fluctuations around the carrying capacity, and the predicted MTE is essentially based on an extrapolation from these typical fluctuations to large ones. One notable exception to the failure of the FP approximation occurs when the population is close to the extinction threshold as predicted by the deterministic rate equation [36, 39]. Thus, the classical results [27-29], which are based on the FP approximation, are correct only for small values of the growth rate parameter $r$.

Interestingly, the difficulty of characterizing extinction mathematically is similar to the difficulty of studying extinction empirically: in empirical research, direct observations of local extinctions are seldom so numerous that the dependence of the extinction risk on the underlying factors can be statistically analyzed (but see [52, 53] for counter examples). Thus, indirect measures, such as the amplitude of population fluctuations, are often used as proxies for the extinction risk [50, 54]. Similarly to the FP approximation, this approach is also based on an extrapolation, i.e. on the assumption that extinction will be caused by similar though somewhat stronger fluctuations than those observed empirically.

In contrast to the FP approximation, the Wentzel-Kramers-Brillouin (WKB) approximation [55], first employed in the context of the master equation in Refs. [56-59], is ideally suited for rare event statistics in large populations, as it accurately predicts quantities such as the MTE and the most likely pathway to extinction [36, 60-63] (Box 3). Both the FP and WKB approximations replace the master equation with an analytically tractable equation, but in the case of the WKB approach, the approximation is mathematically controlled. The WKB approach has gained much recent attention in the physics literature [36, 45, 60-62, 64, 65] but is still to make its headway into ecological modeling.

Both the FP and the WKB approximations treat the scaled population size $n/K$ as a continuous variable, and therefore they are invalid for a very small population size $n$. The WKB approximation, however, can be combined with a small-$n$ approximation [36, 45] to obtain accurate results over the entire state-space (Fig. 1). Small-$n$ approximations are usually based on the observation that density dependence is negligible if the population size is small, in which case a linearised master equation can be exactly solved by recursion.

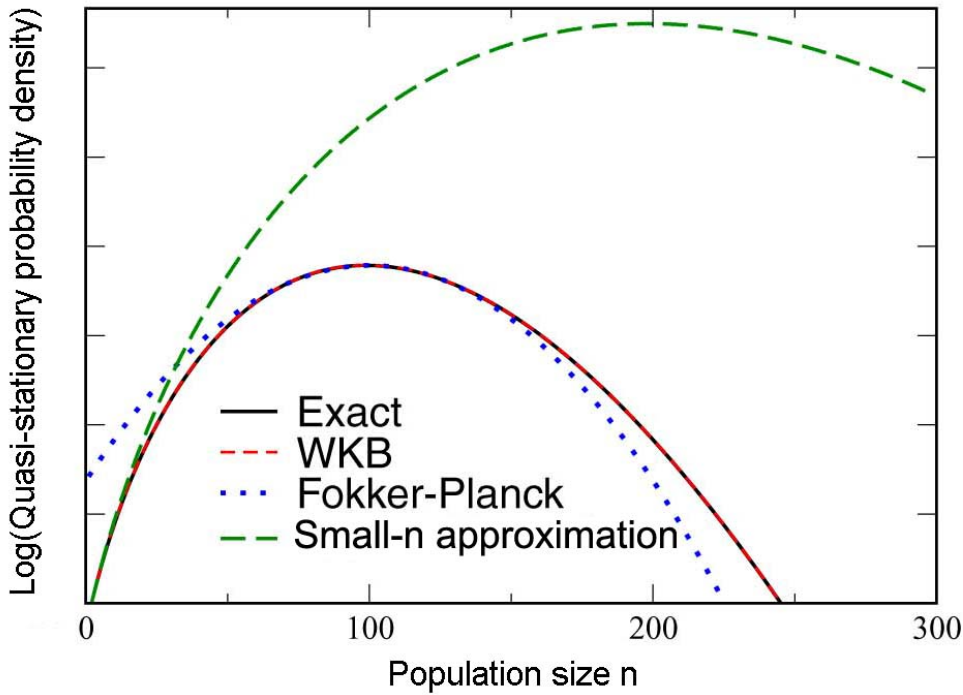

**Figure 1.** The domains of validity for mathematical approximations used to analyze stochastic population models. The FP approximation is valid in the vicinity of the carrying capacity ($K = 100$ in this case), but not for large fluctuations of the population size. As suggested by its name, the small-$n$ approximation is valid only for the smallest population sizes. The WKB approximation very accurately mimics the true distribution for the entire range of population sizes. Modified from [36], to which the reader is referred for the model description and parameter values.

This section would not be complete without mentioning the earlier works in the ecological literature that used the WKB approximation: for a calculation of the stationary probability distribution of a population that does not exhibit extinction [58], and for a calculation of the MTE of a population on the way to extinction, in the case of weak environmental and demographic noises [66, 67]. Unfortunately, the pioneering works [66, 67] employed the WKB method in conjunction with the FP approximation, and as a consequence, their results for the MTE were inaccurate.

## Birth-death processes in a stable environment

Equipped with the mathematical methods, we next review the insight that stochastic population models have brought to the understanding of extinction processes. We start from the simplest case of a birth-death process in a stable environment. Depending on the species life-cycle, it might be natural to formulate population models either in discrete or in continuous time [1, 37]. In the latter case, a large class of models can be described in the framework of Markov processes, examples being given in Box 2. In the deterministic rate equation (Box 3), all the single-step models of Box 2 follow logistic growth, with population size $n$ evolving as

$$\frac{dn(t)}{dt} = rn(t)(1 - n(t)/K). \quad (\text{Eq. 3})$$

Here $r$ is the growth rate of the population at low population density, and $K$ is the carrying capacity that determines the equilibrium population size. The growth rate parameter can be represented as $r = \delta(R_0 - 1)$, where $\delta$ denotes the death rate at low density. The basic reproductive ratio $R_0$ is the average number of offspring produced by a single individual in the absence of intraspecific competition. Density dependence can increase the death rate, decrease the birth rate, or act simultaneously on both processes (Box 2). Density-dependence is usually modeled through a system size parameter $N$, which in our example models is related to the carrying capacity as $K = N(1 - 1/R_0)$ (Box 2).

The prediction of the deterministic approximation (Eq. 3) is that for $R_0 > 1$ (or equivalently $r > 0$), population size approaches the carrying capacity $K$ exponentially in time and persists there for an indefinite time. For $R_0 < 1$, population size decreases to extinction exponentially in time. At the threshold, $R_0 = 1$, population size decreases to extinction as $n(t) = n_0(1 + \delta R_0 n_0 t/N)^{-1}$: much slower than for $R_0 < 1$.

In the full stochastic model, the extinction state is an absorbing state, meaning that the population cannot recover once its size becomes zero. Monte-Carlo simulations above the extinction threshold ($R_0 > 1$) can give the impression of an indefinite persistence (Fig. Ia in Box 2). However, the system eventually goes extinct with probability one [38], though the mean time until extinction is exponentially long in $K$ (Box 2).

Close to the deterministic extinction threshold, with $R_0$ only marginally greater than one, all models in this class predict that the rate of exponential scaling in Eq. 1 is proportional to the population growth rate $r$ [45]. This result also coincides with the classical results, which is expected, as the FP equation is valid close to the extinction threshold [39, 45]. An exponential scaling with the carrying capacity also holds for discrete-time models defined as Markov chains [68], but such models are harder to analyze than their continuous-time counterparts, and mathematically rigorous large-$N$ approximations are generally not available.

In the threshold case $R_0 = 1$ (illustrated by Fig Ib in Box 2), the MTE scales with the square-root of the system size parameter $N$, and the formula $\text{MTE} = (\frac{\pi}{2})^{\frac{3}{2}} \sqrt{N}/\delta + \ln n(0)$ holds in the large-$N$ limit for all single-step models of Box 2 [39]. For $R_0 < 1$, the MTE predicted by the stochastic model coincides with that predicted by the deterministic model, giving (for large $N$) $\text{MTE} = \ln n(0)/(\delta(1 - R_0))$ [39].

## Effects of environmental variability

The message from the stable environment models is that the MTE scales exponentially with the carrying capacity, and thus large populations practically never go extinct during ecological timescales. The main reason for the discrepancy between this prediction and reality is that real populations are also exposed to deleterious processes other than demographic stochasticity. Adding any kind of stochasticity into a population model usually increases the amplitude of population fluctuations and thus extinction risk, but the exact nature of the stochastic process can make a major difference (Fig. 2).

Probably the most important single source of stochasticity affecting population viability is variability in the environmental conditions. Environmental stochasticity can be modeled by assuming that the birth and death rates vary randomly in time [27-29, 49, 69-72], or by assuming that occasional catastrophic events lead to large perturbations in either the birth and death rates [63] or directly in the population size [28, 46, 73]. Environmental noise is filtered through the demographic population processes, and induces non-trivial fluctuations in the population dynamics [74, 75].

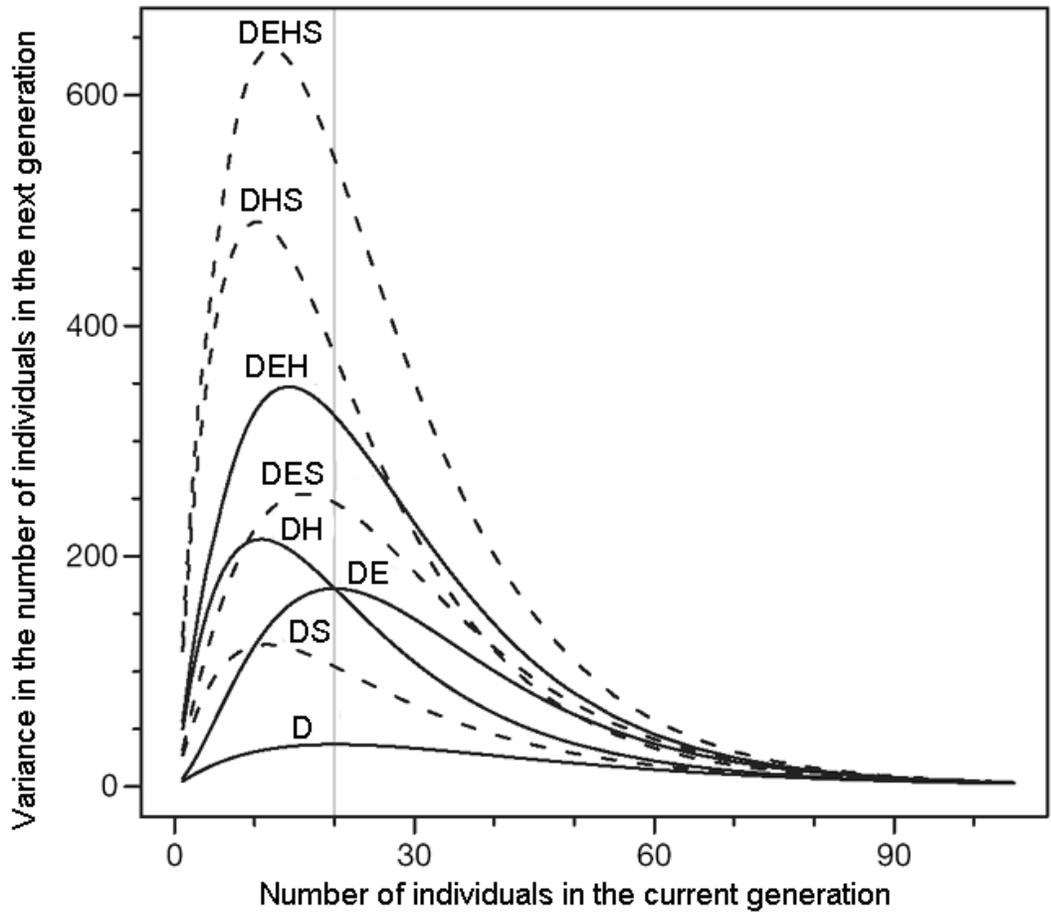

**Figure 2.** The amplitude of population fluctuations and thus the extinction risk strongly depend on the underlying processes generating stochasticity. Each line corresponds to a variant of the stochastic Ricker model. All models include demographic stochasticity (D), and a combination of environmental stochasticity (E) affecting the birth rates, demographic heterogeneity (H) causing variation in birth rates among the individuals, and the stochasticity associated with sex determination (S). The vertical line at $n = 20$ marks the carrying capacity. Modified from [5].

In some models, the demographic stochasticity is neglected altogether by treating the population size as a continuum [71, 72], so that extinction needs to be defined as the population size reaching a prescribed (low) threshold value [72]. Here we focus on models which combine both demographic and environmental stochasticity and thus lead to true extinctions.

The classical papers [27-29] considered variants of the canonical model (Box 2) under the assumption that environmental variations are temporally uncorrelated (white noise) and normally distributed (Gaussian noise). For strong environmental stochasticity they derived the power-law scaling of Eq. 2 using the FP approximation, so this result is well established only when the population is close to the extinction threshold. The power-law scaling has been verified numerically [49, 70] and reproduced, in the same parameter regime, by the WKB method [76]. To our knowledge, the case of a high growth rate is still an open mathematical problem.

Much recent emphasis has been put on the effect of the color of the environmental noise [1, 12, 31-35, 69, 74-76]. Intuitively, slow environmental variations are expected to increase the extinction risk compared to rapid fluctuations. This is because the extinction risk can be strongly elevated by a consecutive series of years with adverse conditions [34, 77], an observation made already by Foley [29]. Theoretical studies, however, have produced conflicting results even when predicting the sign of the effect [77], depending on the time-frame for measuring extinction [34], on interactions between the environmental noise and demographic processes [31, 34, 35, 74], and on the time scale at which the amplitude of environmental noise is measured [32].

Most studies of extinction risks in colored environments are based on Monte-Carlo simulations, analytical results being scarce [69]. Recently, the WKB approach has been used to study extinction of populations affected by a large perturbation [63] or a continuous noise-modulation of the birth and death rates [76]. In the latter case, the MTE depends on the strength of the noise and on its autocorrelation time (Box 4). With weak noise, the model still predicts an exponential dependence on the carrying capacity (Eq. 1), but with a smaller parameter $b$ than with demographic stochasticity alone. If the noise is sufficiently strong and short-correlated, the power-law scaling arises. If the noise has a sufficiently long autocorrelation time, the extinction risk becomes independent of the carrying capacity (Box 4).

## Allee effects

The models considered above assume that the per-capita growth rate is highest when population density is low, and they thus ignore many problems faced by small populations, such as the difficulty of finding mates, increased vulnerability to predation [78], inbreeding depression [7], or more generally Allee effects [1, 20, 79, 80]. An Allee effect can modify the deterministic rate equation (Eq. 3) in such a way that the population growth rate becomes negative below a critical population size $n_c < K$. Then, an unstable fixed point arises at density $n = n_c$ in addition to a stable equilibrium at the carrying capacity $K$ [20]. As the extinction state $n = 0$ also becomes a stable equilibrium, the model exhibits bistability. The critical density $n_c$ thus determines, in the deterministic model, the minimum initial population size required for a successful establishment of a population [20].

The consequences of Allee effects have been analyzed extensively with deterministic models [20, 79] and Monte-Carlo simulations [79, 81], but very few analytical results are available for stochastic models [79]. Dennis [82] used the diffusion approximation to show that extinction in a stochastic model with an Allee effect strongly depends on the initial population size in the vicinity of the critical density $n_c$.

The WKB approach has been recently applied to models with bistability [45, 65]. Therefore, the general mathematical machinery required for the analysis of stochastic models with Allee effects is now available [45], and we expect biological applications in this area to emerge. In ecological research, the influence of an Allee effect has sometimes been modeled non-mechanistically by assuming that the population goes extinct if its size hits the critical size $n_c$ [1]. The results from the WKB approach give justification for this intuitive scenario, as the leading term in the MTE is determined by the demographic-noise -induced transition from the carrying capacity $K$ to the critical population size $n_c$ (Box 3) [45].

## Models with more than one entity type

Models with more than one entity type can be used to extend the biological context in various ways. Examples include demographic heterogeneity [5] due to e.g. age [83] or sex [5, 33] structure, a set of local populations connected by dispersal [26, 69], and the dynamics of interacting species [20].

The analysis of large fluctuations and population extinction in models with multiple entity types is a hot area of research in physics [61, 62, 84-87], especially in the context of disease dynamics.

Quantitative studies on the mean time to extinction of diseases, caused for example by stochasticity of the disease transmission and recovery, were initiated in the pioneering works of Bartlett [88]. Archetypical epidemiological models include the susceptible-infected and susceptible-infected-recovered models [62, 64, 89-91], and the susceptible-infected-susceptible model [61, 84, 87] with population turnover. In these epidemiological models, the mean time to disease extinction is a direct analogue of the MTE in models of ecology.

The mathematical challenges of finding the mean time to disease extinction increase with the number of entity types in every method used. The FP and WKB approaches bring about high-dimensional problems that are generally analytically intractable, except for special cases, e.g. in the parameter regime close to the extinction threshold [61, 62, 85]. In cases where analytical solutions are not available, numerical solutions based on the WKB approach can provide sharp insights into large fluctuations and most likely pathways to disease extinction. Examples of such calculations for the susceptible-infected model are shown in Box 5.

## Understanding population extinction remains a challenge

Stochastic population models can be used to gain insight into how endogenous and exogenous factors interact in determining the fate of populations and in particular for assessing their extinction risk. In this review, we have focused on a relatively simple class of models, in which mathematical analysis beyond stochastic simulations can be used to provide an analytical insight. In spite of extensive research efforts over the past decades, we argue that there is a lack of a general synthesis that would extend much beyond qualitative predictions of the classical models [27-29]. This is partly because there has been a lack of appropriate mathematical machinery for characterizing extinction events, and thus most results are based on either Monte-Carlo simulations or uncontrolled mathematical approximations. However, approaches suitable for the study of large fluctuations and centered on the WKB approximation have been recently developed by physicists. These approaches make it possible to accurately evaluate the mean time to extinction and to determine the most probable pathway of the population on the way to extinction. We propose that research in theoretical ecology takes advantage of these new methods to reach a new level of understanding of the process of population extinction and other rare events of interest to ecologists.

## Acknowledgements

We thank Sami Ojanen for his help with preparing the figures. This study was supported by the Academy of Finland (Grant no. 124242 to OO), the European Research Council (ERC Starting Grant no. 205905 to OO), the Israel Science Foundation (Grant No. 408/08 to BM), and the US-Israel Binational Science Foundation (Grant No. 2008075 to BM).

## Box 1. What are formulae for the mean time to extinction needed for?

This review focuses on studies that have derived mathematical formulae for the mean time to extinction (MTE). One might argue that there is no need to solve simple population models mathematically, as with greatly increased computer power it is possible to simulate the behavior of much more realistic and complex models. Furthermore, such simulations can be used to predict not only the MTE, but also the full distribution of extinction times or any other statistic of interest. So why bother to derive an explicit formula for the MTE, especially given that doing so can be mathematically challenging (Box 3)?

Let us first ask whether MTE is a meaningful statistic or if it would be more informative to focus on other aspects of the extinction process. The answer turns out to be very simple. Assuming that the initial population size is large enough for the population to avoid a rapid initial extinction, the distribution of extinction times is exponential in almost any kind of population model, including very complex individual-based models [92]. Thus, in this case, the MTE is a sufficient statistic for predicting the full distribution of extinction times. However, if the initial population size is so low that the risk of extinction is initially much greater than after the population has reached the quasi-stationary state, the MTE can give a very misleading picture of the extinction risk [93].

Explicit formulae for the MTE, such as Eqs. 1 and 2, provide insight into how extinction risk depends on model parameters. Such information might be obtained with simulations as well, though it can be computationally very challenging to tabulate the MTE for all relevant parameter combinations. However, the main caveat with the simulation approach is that it is difficult to synthesize the numerical results into a format that can be applied and communicated as easily as an explicit equation. Simple mathematical formulae for the MTE have for example been used as building blocks in more complex models [10, 94]. It would be very difficult to use simulation results for this purpose.

## Box 2. Standard models of stochastic population dynamics

In single-step models, the only allowed transitions are births (transition $n \to n+1$ with rate $\lambda_n$) and deaths (transition $n \to n-1$ with rate $\mu_n$) of single individuals. Archetypal models include the **Verhulst model** [43], in which the per-capita birth rate $\Lambda_n = \lambda_n/n = \delta R_0$ is independent of the density, and the per-capita death rate $M_n = \mu_n/n = \delta(1 + R_0 n/N)$ increases with the density. In the **susceptible-infective-susceptible model**, also called the stochastic logistic model [40-42, 44], density dependence operates on births only, $\Lambda_n = \delta R_0(1 - n/N)$ and $M_n = \delta$. In the **symmetric logistic model** [27, 37], density dependence operates on both births and deaths, $\Lambda_n = \delta R_0(1 - n/2N)$ and $M_n = \delta(1 + R_0 n/2N)$. We have parameterized the models so that the death rate at low density is $\delta$, the basic reproductive ratio is $R_0$, and the carrying capacity (the equilibrium population size predicted by the deterministic approximation) is $K = N(1 - 1/R_0)$.

Above the extinction threshold ($R_0 > 1$), and starting from sufficiently many individuals to avoid a rapid initial extinction, for example the susceptible-infected-susceptible model yields the exponential scaling law [39-43]

$$\text{MTE} = \frac{R_0}{\delta(R_0 - 1)^2} \sqrt{\frac{2\pi}{N}} \exp\left(\left(\ln R_0 - 1 + \frac{1}{R_0}\right)N\right). \quad \text{(Eq. I)}$$

This result is an example of a mathematically rigorous large-$N$ approximation, which can be obtained using either the special technique for single-step models [39], or the more general WKB method [45]. Eq. I illustrates the nature of the pre-factor $C_1$ in Eq. 1, which can depend on $N$ (or $K$, which is proportional to $N$), but is asymptotically negligible at large $N$, compared to the exponential behavior. The Verhulst model and the symmetric logistic model also lead to an exponential scaling of the MTE with $N$, with somewhat different exponents and pre-factors [45].

Any of the single-step models can be extended to a **birth-death-catastrophe model** [46] by assuming that catastrophes arrive at a given rate and for example wipe out a binomially distributed number of individuals.

Environmental stochasticity can be added in a multitude of ways. As an example, the **environmental-noise-modulated symmetric logistic model** [76] extends the symmetric logistic model by adding the terms $-\xi(t)/2$ and $\xi(t)/2$ to the per-capita birth and death rates, respectively. Ref [76] parameterized the noise $\xi(t)$ in terms of its variance $v_e$ (describing how greatly the environment varies) and autocorrelation time $t_c$ (describing how fast the environment varies).

Treating the population size as a continuous variable, the above population models can be approximated by the Langevin equation

$$\frac{dn}{dt} = \delta(R_0 - 1)n(1 - n/K) + \sigma_d \xi_d(t)\sqrt{n} + \sigma_e \xi_e(t)n. \qquad \text{(Eq. II)}$$

Here the demographic noise $\xi_d$ is white and Gaussian with zero mean and variance one (Box 3). The environmental noise $\xi_e$ can in general be colored. For the particular case of white environmental noise, Eq. II has been termed the **canonical model** [26, 49].

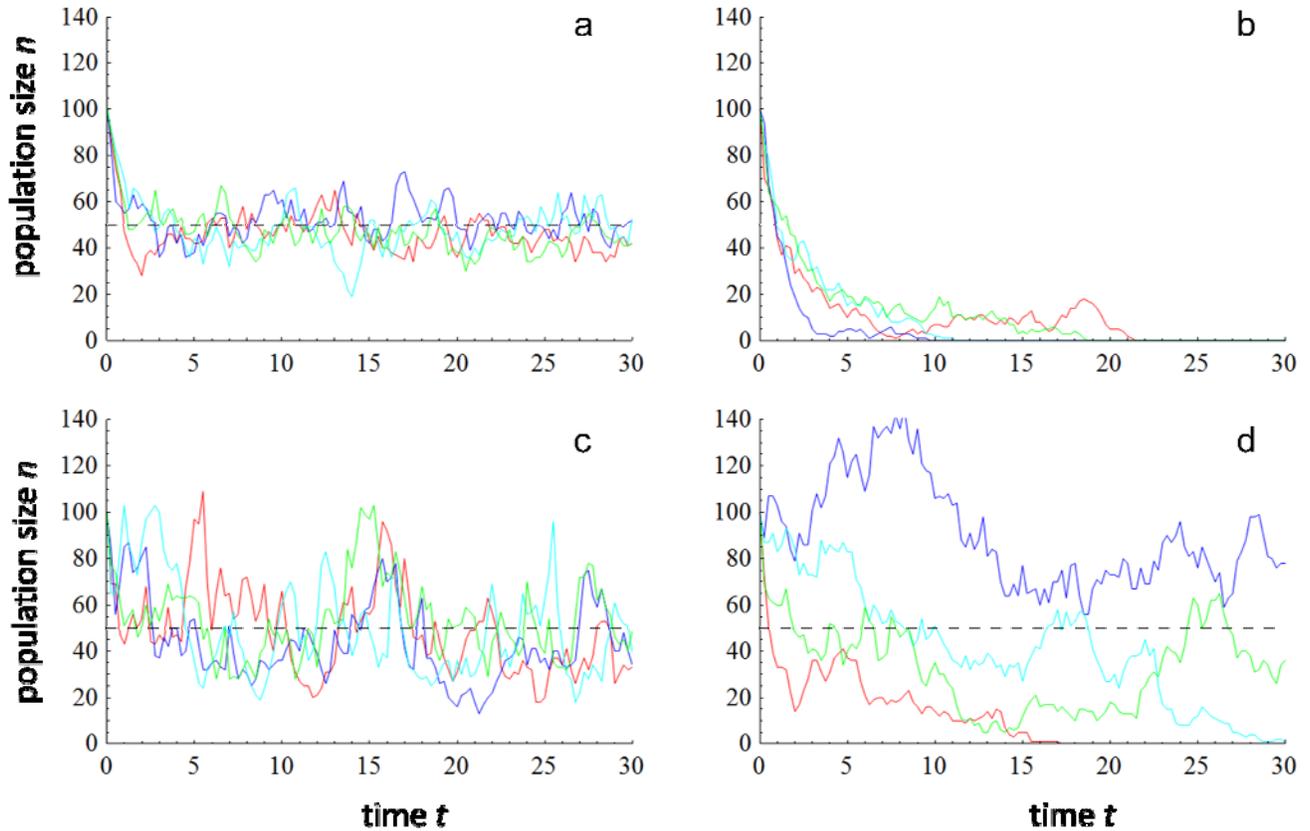

**Figure I.** Extinction risk and variability among replicate simulations depends on the presence and type of environmental noise and parameter regime. The panels show Monte-Carlo simulations of the symmetric logistic model without (**a,b**) and with (**c,d**) environmental noise modulation. In all panels, $N = n(0) = 100$, $\delta = 1$ and $R_0 = 2$, except in (**b**) where $R_0 = 1$. In (**c**), $v = 1$ and $t_c = 1$, in (**d**) $v = 1$ and $t_c = 30$. The dashed lines depict the deterministic carrying capacity $K = N(1 - 1/R_0)$, and the colored lines show four independent realizations.

# Box 3. Master equation and its approximations

The dynamics of any continuous-time population model in a stable environment can be mathematically described by the master equation. We denote by $p_n(t)$ the probability that the population has $n$ individuals at time $t$, and by $\boldsymbol{p}(t)$ the probability distribution of population sizes, i.e. the vector with elements $p_n(t)$. By the theory of Markov processes [38], $\boldsymbol{p}(t)$ evolves in time as

$$\frac{d\boldsymbol{p}(t)}{dt} = \boldsymbol{Q}\boldsymbol{p}(t). \quad \text{(Eq. I)}$$

Here $\boldsymbol{Q}$ is called the transition rate matrix, with elements describing the rates (probabilities per unit time) with which the system moves from the current population size to any other population size. When eventual extinction is certain, the limiting (stationary) distribution $\boldsymbol{p}(\infty)$ is such that $p_0(\infty) = 1$ and $p_n(\infty) = 0$ for $n > 0$. If the time to extinction is sufficiently long, the system approaches a quasi-stationary state, with a time-independent shape distribution $\pi_n$ of population sizes with $n > 0$ [42, 43, 83]. In this state $p_0(t) \cong 1 - \exp(-t/\text{MTE})$ and $p_n(t) \cong \pi_n \exp(-t/\text{MTE})$ for $n > 0$, whereas the time to extinction is exponentially distributed with average equal to MTE [45, 92]. Mathematically, $\pi_n$ is the probability distribution of the population sizes conditional on non-extinction.

The diffusion, or Fokker-Planck approximation, can be obtained by treating the population size $n$ as a continuum, expanding the master equation in a Taylor series, and truncating the expansion at the second order [38]. The resulting partial differential equation has the form

$$\frac{\partial P(n,t)}{\partial t} = -\frac{\partial}{\partial n}[F(n)P(n,t)] + \frac{1}{2}\frac{\partial^2}{\partial n^2}[D(n)P(n,t)]. \quad \text{(Eq. II)}$$

Here $F(n)$ is the right-hand side of the deterministic rate equation (Eq. 3) which, for the single-step models of Box 2, reads $F(n) = \lambda_n - \mu_n$. The diffusion term $D(n) = \lambda_n + \mu_n$ describes the intrinsic noise that arises from demographic stochasticity. The FP equation can be transformed into an equivalent Langevin equation that describes the fate of individual realizations:

$$\frac{dn}{dt} = F(n) + \sqrt{D(n)}\xi_d(t), \quad \text{(Eq. III)}$$

where $\xi_d$ is Gaussian white noise with zero mean and unit variance.

In the WKB-approximation [45, 55, 56, 59, 64], the quasi-stationary solution $\pi_n$ to the master equation is approximated by an exponentiated expansion in powers of $1/K$:

$$\pi_n = \exp(-KS(n/K) - S_1(n/K) - S_2(n/K)/K - \cdots), \quad \text{(Eq. IV)}$$

The term $S(n/K)$, called the action, determines the leading-order behavior of the system, and is sufficient for calculating the coefficient $b$ in the exponential scaling of Eq. 1. Substituting Eq. IV into the master equation (Eq. I) results in a differential equation for the leading term $S(n/K)$. This equation has a form of the Hamilton-Jacobi equation of classical mechanics [95]. The quasi-stationary distribution $\pi_n$ corresponds, in the language of Hamiltonian mechanics, to a phase trajectory with zero energy. The zero-energy trajectory with zero momentum corresponds to the deterministic component $F(n)$, whereas a zero-energy trajectory with a nonzero momentum describes how the rare event of population extinction is most likely to happen (Fig I).

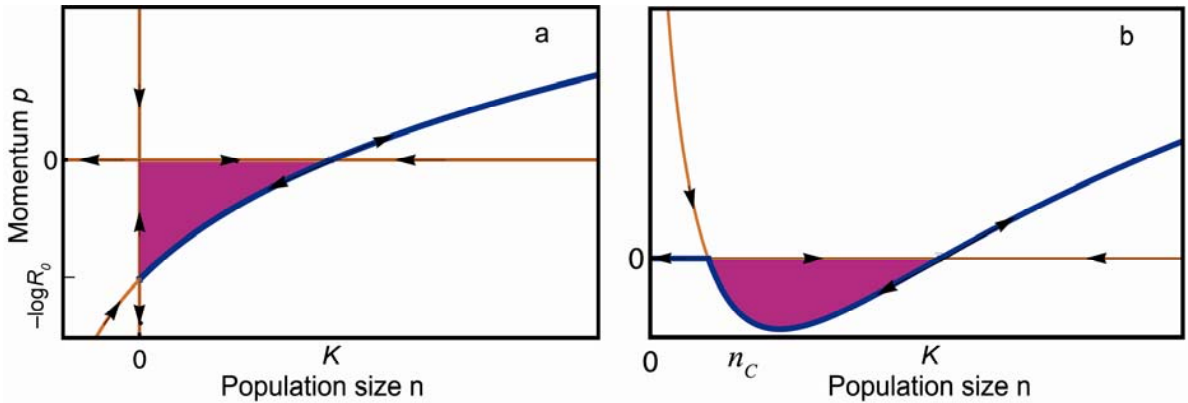

**Figure I.** An illustration of the WKB method for estimating the extinction risk in a model without (**a**) and with (**b**) an Allee effect. The zero-energy trajectories with zero momentum ($p = 0$), shown by the horizontal brown lines, correspond to the deterministic rate equation. The zero-energy trajectories with a nonzero momentum, shown by the blue lines, describe how the rare event of population extinction is most likely to happen. The total action along such a special trajectory, given by the area of the shaded region, is sufficient for calculating the MTE to the leading order. Redrawn from [45].

# Box 4. Pathways to extinction under environmental noise

Here we illustrate the power of the WKB method by characterizing population extinction in the environmental-noise-modulated symmetric logistic model (Box 2) [76]. This analysis is restricted to the parameter regime where the basic reproductive ratio $R_0$ is only slightly above one and the carrying capacity $K$ is large. Without environmental noise, the MTE in this model behaves (up to the pre-factor $C_1$ of Eq. 1) as $\exp(b_0 K)$, where $b_0 = r/(2\delta)$.

The qualitative effect of environmental noise depends on its autocorrelation time $t_c$ compared to the relaxation time $t_r = 1/r$ needed for the population to reach the carrying capacity in the deterministic model. For short-correlated ($t_c \ll t_r$) and weak ($v_e \ll r\delta/K$) noise, the MTE still scales exponentially with $K$, but with a reduced coefficient $b = b_0 - v_e t_c r K^2/(6\delta^2)$ [76]. Thus, a large population will have a negligible extinction risk if the environmental fluctuations are sufficiently mild and if they vary at a short time scale so that the effects of adverse conditions cannot accumulate in time.

Very adverse environmental conditions lead to a quick extinction, but if such conditions are extremely rare, there is a trade-off between the statistical weight of the noise and its impact on population dynamics [34, 76]. For strong ($v_e \gg r\delta/K$) but short-correlated noise, the time to extinction scales as the power-law $(v_e t_c K/\delta)^{r/v_e t_c}$ [27-29, 76]. The WKB method yields the optimal realization of noise which determines the most probable path to extinction, and thus explains where the power-law scaling comes from. In the optimal realization, the noise attains, for some time, the value $\xi(t) = 2r$ leading the system on a deterministic pathway to extinction. The duration of this event is relatively small, $\ln(Kv_e t_c/2\delta)/r$. Thus, if the environmental conditions undergo strong and fast fluctuations, extinctions are mostly caused by catastrophes (short-lived but very severe conditions).

In case of long-correlated noise ($t_c \gg t_r$), the optimal realization of noise, predicted by the WKB method, is such that the adverse conditions continue much longer than the relaxation time of population dynamics. However, they are relatively mild in the sense that the population growth rate stays above the deterministic threshold even during the extinction process.

Ref [76] gives an explicit formula for the MTE for both weak and strong long-correlated noises. In the latter case, the expression for the MTE is especially simple, $\text{MTE} \sim \exp(r^2/2v_e)$. Thus, the WKB method predicts that, while short-correlated noise leads to a power-law scaling, with long-correlated noise MTE becomes (up to a pre-factor) independent of carrying capacity.

# Box 5. Pathways to epidemic outbreak and disease extinction

Here we illustrate how the WKB method can be used to quantify extinction in models with two or more entity types. We consider the susceptible-infected model with population turnover [62, 85, 89-91], with $S$ standing for the number of susceptible and $I$ for the number of infected individuals. A susceptible individual becomes infective at per capita rate $(\beta/N)I$, where $\beta$ is the infection rate, and $N$ scales as the size of the entire population. Susceptible individuals die at rate $\mu$, infected individuals die at rate $\mu_I$, and new susceptibles are brought into the population at rate $\mu N$.

In the deterministic version of the susceptible-infected model, the infection rate needs to exceed the critical threshold $\beta > \mu_I$ for the disease to persist [90]. Above the threshold, the deterministic model has a stable fixed point, with a constant proportion of the individuals remaining infected (Figure I). In the full stochastic model, demographic stochasticity causes the ultimate extinction of the disease for any parameter values. Applying the WKB approximation, one can calculate numerically the mean time to disease extinction, the probability of the disease going extinct immediately after the first epidemic outbreak, and the most likely paths of the system to disease extinction [62, 85] (Fig. I). Analytical results can be obtained in special cases, for example, close to the critical threshold. Similar to problems with one entity type, the mean time to disease extinction grows exponentially with the parameter $N$ [62, 85].

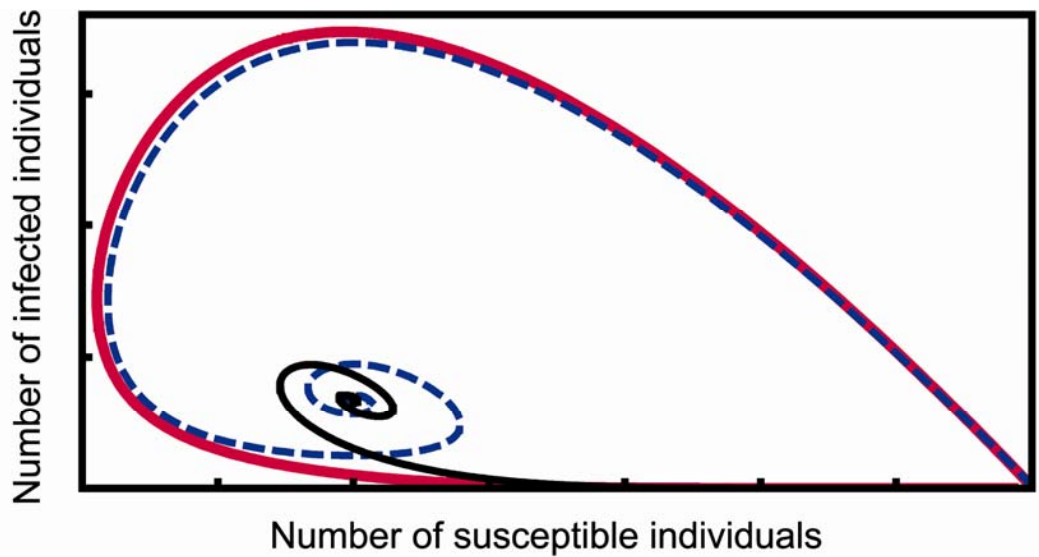

**Figure I.** Most likely paths to an epidemic outbreak and to disease extinction as predicted by the WKB method. The dashed blue line shows the prediction of the deterministic susceptible-infected model after a few infected individuals have been introduced into the host population. Here, after the first epidemic outbreak, the number of infected individuals oscillates towards a fixed point, where the disease persists indefinitely. In the stochastic model, the introduction of a few infected individuals can lead either to the deterministic pathway, or to two other major possibilities: a rapid extinction of the disease after the introduction (not shown), or the extinction of the disease after the first epidemic outbreak (red line). Even if the disease reaches the endemic state (the fixed point of the deterministic model), it eventually goes extinct in the stochastic model, the black line showing the most likely path for this to happen. Redrawn from ref [85].

# Glossary

**Allee effect:** Refers to a variety of processes that reduce the per-capita growth rate at small population density.

**Bistability:** Presence of two different stable populations sizes, as predicted by deterministic rate equation.

**Colored noise:** Refers to temporal autocorrelation structure of the noise associated with environmental stochasticity. White noise is uncorrelated in time, so that future environmental conditions do not depend on earlier environmental conditions. Different colors refer to situations where the environmental conditions vary slower (red or pink noises, positive correlation) or faster (blue noise, negative correlation) than for the white noise.

**Demographic heterogeneity:** Variation in the intrinsic birth and death rates among individuals.

**Demographic stochasticity:** Random variation in the number of births and deaths in a population caused by the discrete nature of individuals and stochastic character of these processes. Present even if all individuals have identical birth and death rates.

**Diffusion approximation:** See Fokker-Planck approximation.

**Entity type:** Different entity types can refer to different species, different local populations, different age classes, or any other structure that requires differentiating between classes of individuals in a population model.

**Environmental stochasticity:** Irregular variation in environmental conditions that affect the birth and death rates of a population.

**Extinction threshold:** In deterministic models, the population either goes eventually extinct or persists for an indefinite time. The part of the parameter space that separates these two qualitative behaviors comprises the extinction threshold.

**Fokker-Planck (FP) approximation:** Transforms the master equation into a simpler but approximate partial differential equation (Box 3).

**Markov process:** A stochastic process without memory, i.e. a process in which the transition rates depend only on the current state of the system.

**Master equation:** An exact equation for population dynamics in Markov process models. Describes how the probability distribution of population sizes evolves in time (Box 3).

**Monte Carlo simulation:** Produced by a computational algorithm that repeatedly utilizes a random number generator to construct realizations of a stochastic process (Fig I in Box 2).

**Probability distribution:** Defines the state of the system, i.e. the probability $p_n(t)$ that the population consists of $n$ individuals at time $t$.

**Quasi-stationary probability distribution:** The limit of the probability distribution at large time, conditional on the population being not yet extinct. In models that eventually lead to extinction, the stationary probability distribution describes the state where the population is extinct, whereas the quasi-stationary distribution describes the shape of the distribution of the population sizes long after an initial transient but before extinction (Box 3).

**Single-step and multi-step models:** In single-step models, the population increases or decreases by one individual at a time, whereas in multi-step models multiple simultaneous births or deaths are allowed for (Box 2).

**Stationary probability distribution:** The limit of the probability distribution at large time.

**System size parameter $N$:** A model parameter that describes the population size at which density dependence has a substantial effect (for example, death rate doubled from the density-independent level, birth rate equals zero, or so on). If the population persists in the deterministic approximation, $N$ is proportional to the equilibrium population size $K$ (the carrying capacity), but the proportionality constant is somewhat arbitrary and depends on how $N$ is defined in the particular model.

**Wentzel-Kramers-Brillouin (WKB) approximation:** A mathematical tool for transforming the master equation into a simpler but approximate set of ordinary differential equations (Box 3).

65 Escudero, C. and Kamenev, A. (2009) Switching rates of multistep reactions. *Physical Review E* 79, 041149

66 Tier, C. and Hanson, F.B. (1981) Persistence in density dependent stochastic populations. *Math. Biosci.* 53, 89-117

67 Hanson, F.B. and Tier, C. (1981) An asymptotic solution of the first passage problem for singular diffusion in population biology. *SIAM Journal on Applied Mathematics* 40, 113-132

68 Klokov, S.A. (2009) Upper estimates of the mean extinction time of a population with a constant carrying capacity. *Mathematical Population Studies* 16, 221-230

69 Hill, M.F.*, et al.* (2002) The effects of small dispersal rates on extinction times in structured metapopulation models. *Am. Nat.* 160, 389-402

70 Hakoyama, H.*, et al.* (2000) Comparing risk factors for population extinction. *J. Theor. Biol.* 204, 327-336

71 Ji, C.Y.*, et al.* (2009) Analysis of a predator-prey model with modified Leslie-Gower and Holling-type II schemes with stochastic perturbation. *Journal of Mathematical Analysis and Applications* 359, 482-498

72 Braumann, C.A. (2008) Growth and extinction of populations in randomly varying environments. *Computers & Mathematics with Applications* 56, 631-644

73 Cairns, B.J. (2009) Evaluating the expected time to population extinction with semi-stochastic models. *Mathematical Population Studies* 16, 199-220

74 Ruokolainen, L.*, et al.* (2009) Ecological and evolutionary dynamics under coloured environmental variation. *Trends Ecol. Evol.* 24, 555-563
**40**